\documentclass[12pt]{article}
\makeatletter
\newcommand{\singlespacing}{\let\CS=\@currsize\renewcommand{\baselinestretch}{1}\tiny\CS}
\usepackage{epsfig}

\usepackage{graphicx}
\usepackage{verbatim}   
\usepackage{color}      
\usepackage{subfigure}  
\usepackage{amsfonts}

\begin{document}
\baselineskip=24pt
\parskip = 10pt
\def \qed {\hfill \vrule height7pt width 5pt depth 0pt}
\newcommand{\ve}[1]{\mbox{\boldmath$#1$}}
\newcommand{\IR}{\mbox{$I\!\!R$}}
\newcommand{\1}{\Rightarrow}
\newcommand{\bs}{\baselineskip}
\newcommand{\esp}{\end{sloppypar}}
\newcommand{\be}{\begin{equation}}
\newcommand{\ee}{\end{equation}}
\newcommand{\beanno}{\begin{eqnarray*}}
\newcommand{\inp}[2]{\left( {#1} ,\,{#2} \right)}
\newcommand{\eeanno}{\end{eqnarray*}}
\newcommand{\bea}{\begin{eqnarray}}
\newcommand{\eea}{\end{eqnarray}}
\newcommand{\ba}{\begin{array}}
\newcommand{\ea}{\end{array}}
\newcommand{\nno}{\nonumber}
\newcommand{\dou}{\partial}
\newcommand{\bc}{\begin{center}}
\newcommand{\ec}{\end{center}}
\newcommand{\2}{\subseteq}
\newcommand{\cl}{\centerline}
\newcommand{\ds}{\displaystyle}
\newcommand{\mr}{\mathbb{R}}
\newcommand{\mn}{\mathbb{N}}
\def\refhg{\hangindent=20pt\hangafter=1}
\def\refmark{\par\vskip 2.50mm\noindent\refhg}

\title{\sc On a General Class of Discrete Bivariate Distributions}

\author{\sc Debasis Kundu \footnote{Department of Mathematics and Statistics, Indian Institute of
Technology Kanpur, Pin 208016, India.  E-mail:
kundu@iitk.ac.in, Phone no. 91-512-2597141, Fax no. 91-512-2597500.}}

\date{}
\maketitle

\begin{abstract}

In this paper we develop a very general class of bivariate discrete distributions.  The basic idea is very simple.  The 
marginals are obtained by taking the random geometric sum of a baseline distribution function.  The proposed class of 
distributions is a very flexible class of distributions in the sense the marginals can take variety of shapes.  It can be 
multimodal as well as heavy tailed also.  It can be both over dispersed as well as under dispersed.  Moreover, the correlation 
can be of a wide range.  We discuss different properties 
of the proposes class of bivariate distributions.  The proposed distribution has some interesting physical interpretations 
also.  Further, we consider
two specific base line distributions namely; Poisson and negative binomial distributions for illustrative purposes.  Both 
of them are infinitely divisible.  The 
maximum likelihood estimators of the unknown parameters cannot be obtained in closed form.  They can be obtained by solving
three and five dimensional non-linear optimizations problems, respectively.  To avoid that we propose to use the method of 
moment estimators and they can be obtained quite conveniently.  The analyses of two real data sets have been 
performed to show the effectiveness of the proposed class of models.  Finally, we discuss some open problems and conclude 
the paper.

\end{abstract}



\section{\sc Introduction}

Bivariate lifetime distributions are mainly used to analyze the marginals and also to study the dependence structure of the two 
marginals.  An extensive amount of work has been done to propose different continuous bivariate distributions, analyzing 
their properties and develop different estimation procedures of the unknown parameters.  Some of the early references on 
bivariate distributions are Gumbel \cite{Gumbel:1960}, Freund \cite{Freund:1961}, Marshall and Olkin \cite{MO:1967}, 
Block and Basu \cite{BB:1974} and for some of the recent work one is referred to Balakrishnan and Lai \cite{BL:2009}, 
Kundu and Gupta \cite{KG:2009}, Kundu et al. \cite{KBJ:2010}, Lee 
and Cha \cite{LC:2014}, Sankaran and Kundu \cite{SK:2014}  and the references cited therein.

It may be mentioned that that in many practical situations, even if the data are discrete, they are modelled by continuous 
distributions mainly due to analytical tractability.  But, in many practical situations, the discrete data occur quite 
naturally.  Sometimes, it is impossible to measure life length of a device on a continuous scale.  In many practical situations 
lifetimes are recorded on a discrete scale.  For example, the on/off switching devices, the number of accidents, to and fro 
motion of spring devices etc. are purely discrete in nature.  Bivariate discrete data also occur quite naturally in practice.
For example, the number of goals scored in a football match by two competing teams or the number of insurance claims for two 
different causes is purely discrete in nature.

An extensive amount of work has been done introducing different bivariate discrete distributions, analyzing their properties 
and developing different estimation procedures.  Special attention has been paid on bivariate geometric distributions and 
bivariate Poisson distributions, see for 
example Kocherlakota and Kocherlakota \cite{KK:1992}, Kocherlakota \cite{Koch:1995}, Basu and Dhar \cite{BD:1995}, 
Kumar \cite{Kumar:2008}, Kemp \cite{Kemp:2013}, Lee and Cha \cite{LC:2014}, Nekoukhou and Kundu \cite{NK:2017}, 
Kundu and Nekoukhou \cite{KN:2018} and see the references cited therein.

Recently, Lee and Cha \cite{LC:2015} proposed two classes of discrete bivariate distributions and discussed their properties.
Their idea is based on the minimum and maximum of two independent non-identical distributed random variables.  The idea is quite 
simple, and it produces different unimodal shapes of bivariate discrete distributions.  Unfortunately, because of the 
non-identical distributions, the joint probability mass functions (PMFs) or the marginal PMFs may not be in a convenient form.
It makes it difficult to compute the estimates of the unknown parameters, and to derive different properties.  Moreover, the 
marginals produced by the method of Lee and Chao \cite{LC:2015} cannot have multimodal or heavy tailed property, see also  
Nekoukhou and Kundu \cite{NK:2017} in this respect.

The main aim of this paper is two fold.  First of all we develop a very flexible class of discrete bivariate distributions.  The
proposed discrete bivariate distribution has a very convenient joint probability generating function (PGF), hence the joint 
PMF can be obtained in a convenient form.  The main idea is quite simple.  We consider geometric random sum of independent 
identically distributed (i.i.d.) base line random variables.  The idea is not new.  It has been used in case of continuous 
random variables by several authors.  For example, Chahkandi and Ganjali \cite{CG:2009} used this idea when the base 
distribution is exponential and Barreto-Souza \cite{Barr:2012} extended the results in case of gamma distribution.
The author \cite{Kundu:2014} considered the case when the base distribution is univariate normal and the results have 
been recently generalized for multivariate normal base distribution by the author \cite{Kundu:2017}.  For some of the related
literature, interested readers are referred to Kuzobowski et al. \cite{KPP:2008} and \cite{KPQ:2011}.  Although, the above
method seems to be a powerful method and it has been successfully used for several continuous distributions, no attempt has
been made in case of discrete distributions.  This is an attempt towards that direction.

The main advantage of the proposed method is that it produces a very flexible class of distribution functions.  In this case the
marginals PMFs can be unimodal or multimodal, moreover they can be heavy tailed also.  Since, many well known discrete 
distributions like Poisson, geometric and negative binomial have convolution properties, several interesting properties of the 
proposed distributions can be established.  The proposed model has some interesting physical interpretations also.  
We derive different properties of the proposed class of distributions in general 
and discuss in details two special cases.  It is observed that the marginals can be both under dispersed as well as 
over dispersed also, the correlation can be of a  wide range, hence it can be used quite conveniently for different data analysis 
purposes.

Estimation of the unknown parameters is always an important problem in any data analysis.  In this case it is observed that 
the maximum likelihood estimators (MLEs) of the unknown parameters cannot be obtained in closed forms.  It involves solving 
higher dimensional optimization problems.  To avoid that we propose to use method of moment estimators (MMEs).
It is observed that when the base line 
distribution is Poisson or negative binomial the MMEs can be obtained quite conveniently.  We analyze two real data sets for 
illustrative purposes.

The rest of the paper is organized as follows.  In Section 2, we provide some motivations of the proposed model.  The model 
formulation and some basic properties are discussed in Section 3.  In Sections 4 and 5, we discuss two special cases.  
The analyses of two data sets are presented in Section 6, and 
finally in Section 7 we conclude the paper.

\section{\sc Motivations}

To motivate our proposed model we will start with an example.  First let us consider the traditional construction of the 
bivariate Poisson distribution.  It is based on a trivariate reduction technique and it can be described as follows.  Suppose
$X_1$, $X_2$ and $X_3$ are three independent Poisson random variables with mean $\lambda_1$, $\lambda_2$ and $\lambda_3$, 
respectively.  Consider a new bivariate random variable $(Y_1, Y_2)$, where
$$
Y_1 = X_1 + X_3 \ \ \ \ \hbox{and} \ \ \ \ Y_2 = X_2 + X_3.
$$
It was originally proposed by Holgate \cite{Hol:1964}, see also Campbell \cite{Camp:1934} in this respect. The joint PMF 
of $Y_1$ and $Y_2$ can be obtained as
$$
P(Y_1 = i, Y_2 = j) = e^{-(\lambda_1 + \lambda_2 + \lambda_3)} \sum_{k=0}^{\min\{i,j\}} \frac{\lambda_1^{i-k} \lambda_2^{j-k} \lambda_3^k}
{(i-k)!(j-k)!k!}; \ \ \ \ i,j \in \mathbb{N}_0, 
$$
where $\mathbb{N}_0 = \{0, 1, 2, \ldots\}$.  In this case the marginals will follow Poisson distribution.  Clearly, this is 
an advantage of this model.  But the marginal PMFs cannot be heavy tailed or have multimodal properties.  Due to 
presence of the summation in the joint PMF computing the MLEs become difficult.  Moreover, it may not be very easy to 
generalize it for other bivariate distributions.

Recently Lee and Cha \cite{LC:2015} introduced two very general methods to generate bivariate discrete distribution functions.
The methods can be briefly described as follows.  Let $X_1, X_2$ and $X_3$ be same as defined before.

\noindent {\sc Lee-Cha Method 1:}
$$
U_1 = \max\{X_1, X_3\} \ \ \ \ \hbox{and} \ \ \ \  U_2 = \max\{X_1, X_3\}.
$$
\noindent {\sc Lee-Cha Method 2:}
$$
V_1 = \min\{X_1, X_3\} \ \ \ \ \hbox{and} \ \ \ \  V_2 = \min\{X_1, X_3\}.
$$
In both cases the generated bivariate distributions can be quite flexible.  The joint PMF and the marginals PMFs can be 
of different shapes.  Unfortunately, in this case also it has been observed, see Nekoukhou and Kundu \cite{NK:2017}, 
that the marginals may not be in a very convenient form.  Moreover, in this case also in case of standard discrete distributions
like Poisson, geometric or binomial, the marginal PMFs cannot be multimodal or heavy tailed.  Most of the other bivariate 
distributions as proposed by Kumar \cite{Kumar:2008} and Piperigou and Papageorgiou \cite{PP:2003} have similar 
deficiencies.

The main aim of this paper is to propose a class of discrete bivariate distributions which has convenient joint PMF and 
marginal PMFs.  The joint PMF and the marginal PMFs should be flexible enough and the marginals PMFs can have heavy tailed 
and multimodal shapes.

\section{\sc A general Class of Discrete Bivariate Distributions}

In this section we develop a general class of discrete bivariate distributions and discuss its different properties.  Two 
special cases will be discussed in the subsequent sections.

Suppose $U_1, U_2, \ldots$, are independent identically distributed (i.i.d.) random variables with the probability mass 
function (PMF) $f_1(x)$, for $x \in \mathbb{N}_0 = \{0, 1, 2, \ldots\}$.  Further, $V_1, V_2, \ldots$, are i.i.d. 
random variables with the PMF $f_2(x)$, for $x \in \mathbb{N}_0$, and $N$ is a geometric random variable with the PMF
\be
P(N=n) = \theta(1-\theta)^{n-1}; \ \ \ \ n \in \mathbb{N} = \{1, 2, \ldots\},
\ee
for $0 < \theta < 1$.
All the random variables are independently distributed.  We define the bivariate random variable $(X,Y)$ as follows
$$
X = \sum_{i=1}^N U_i \ \ \ \ \hbox{and} \ \ \ \ Y = \sum_{i=1}^N V_i.
$$
The above bivariate random variable $(X, Y)$ has the following physical interpretations.

\noindent {\sc Accident Model:} Suppose $N$ is the number of accidents that took place in a given place during a fixed period of 
time.  Let $U_i$ and $V_i$ denote the number of deaths and number of seriously injured individuals, respectively,  due to the 
$i$-th accident for $i = 0, 1, \ldots, N$.  Then 
$X$ and $Y$ denote the total number of deaths and total number of seriously injured individuals during that fixed period of time 
due to accidents in that given place.

\noindent {\sc Soccer Model:} Suppose $N$ denotes the number of soccer games played during a year between Team A and Team B.  Suppose 
$U_i$ and $V_i$ denote the number of goal scored by Team A and Team B, respectively, at the $i$-th game, for $i = 0, 1, \ldots, N$. 
Then  $X$ and $Y$ denote the total number goals scored by Team A and Team B, respectively, against each other in that year.

From now on we will call $f_1(x)$ and $f_2(y)$ as the base line PMFs.  First 
we would like to derive different properties of the bivariate random variable $(X, Y)$ for general base line PMFs.  We use the following notations.
$F_1(x)$ and $F_2(x)$ denote the cumulative distribution functions (CDFs) of $f_1(x)$ and $f_2(x)$, respectively. Moreover, $\phi_1(t)$
and $\phi_2(s)$ denote the moment generating functions (MGFs) of $f_1(x)$ and $f_2(x)$, respectively.  We define $f_1^{(n)}(x)$ and 
$f_2^{(n)}(y)$ as the $n$-fold convolutions of $f_1(x)$ and $f_2(x)$, respectively, i.e.
$$
f_1^{(n)}(x) = P(U_1 + \ldots + U_n = x) \ \ \ \ \hbox{and} \ \ \ 
f_2^{(n)}(y) = P(V_1 + \ldots + V_n = y).
$$
We further define $F_1^{(n)}(x)$ and $F_2^{(n)}(y)$ are the CDFs correspond to the PMFs $f_1^{(n)}(x)$ and $f_2^{(n)}(y)$, respectively.
The joint PMF of $X$ and $Y$ for $m = 0, 1, \ldots$ and $n = 0, 1, \ldots$, can be obtained as follows:
\bea
P \left (X = m, Y = n \right ) & = & P \left ( \sum_{i=1}^N U_i = m, \sum_{j=1}^N V_j = n \right )  \nonumber \\
& = & \sum_{k=1}^{\infty} P \left ( \sum_{i=1}^N U_i = m, \sum_{j=1}^N V_j = n \Bigg|  N = k \right ) P(N = k)  \nonumber \\
& = & \sum_{k=1}^{\infty} P \left ( \sum_{i=1}^k U_i = m, \sum_{j=1}^k V_j = n \right ) P(N = k)  \nonumber \\
& = & p \sum_{k=1}^{\infty} (1-p)^{k-1} f_1^{(k)}(m) f_2^{(k)}(n).  \label{jpmf}
\eea
The joint CDF of $X$ and $Y$ for $m = 0, 1, \ldots$ and $n = 0, 1, \ldots$, is
\be
P \left (X \le m, Y \le n \right )  = p \sum_{k=1}^{\infty} (1-p)^{k-1} F_1^{(k)}(m) F_2^{(k)}(n).  \label{jcdf}
\ee
From the joint PMF of $X$ and $Y$, we immediately obtain the marginals PMFs of $X$ and $Y$ as
\be
P(X = m) =  p \sum_{k=1}^{\infty} (1-p)^{k-1} f_1^{(k)}(m) \ \ \ \ \hbox{and} \ \ \ \ \ 
P(Y = n) = p \sum_{k=1}^{\infty} (1-p)^{k-1} f_2^{(k)}(n).
\ee
\be
P(X \le m) =  p \sum_{k=1}^{\infty} (1-p)^{k-1} F_1^{(k)}(m) \ \ \ \ \hbox{and} \ \ \ \ \ 
P(Y \le n) = p \sum_{k=1}^{\infty} (1-p)^{k-1} F_2^{(k)}(n).
\ee
Let us assume that $\phi_1(t)$ and $\phi_2(s)$ exist for $t \in A$ and $s \in B$, where $A, B \subset \mathbb{R}$.  Then for
$(t,s) \in A \times B$, the joint MGF of $X$ and $Y$ is
\bea
\phi_{X,Y}(t,s)  =  E \left ( e^{t X + s Y} \right )  
& = & p \sum_{m=0}^{\infty} \sum_{n=0}^{\infty} \sum_{k=1}^{\infty} e^{tm + s n}(1-p)^{k-1} f_1^{(k)}(m) f_2^{(k)}(n)  \nonumber \\
& = & p \sum_{k=1}^{\infty} (1-p)^{k-1} \sum_{m=0}^{\infty} e^{tm}f_1^{(k)}(m)  \sum_{n=0}^{\infty} e^{sn}f_2^{(k)}(n)  \nonumber \\
& = & p \sum_{k=1}^{\infty} (1-p)^{k-1} \phi_1^k(t) \phi_2^k(s)  \nonumber \\
& = & \frac{p \phi_1(t) \phi_2(s)}{1 - (1-p)\phi_1(t) \phi_2(s)}.   \label{jmgf}
\eea
From the joint MGF of $X$ and $Y$, we obtain for $t \in A$ and $s \in B$,
\be
\phi_X(t) = \frac{p \phi_1(t)}{1 - (1-p)\phi_1(t)} \ \ \ \ \hbox{and} \ \ \ \
\phi_Y(s) = \frac{p \phi_2(s)}{1 - (1-p)\phi_2(s)}.
\ee
Moreover, it is immediate from (\ref{jmgf}) that $X$ and $Y$ are independent if and only if $p$ = 1.  Using the MGFs or otherwise, 
different moments and product moments can be easily obtained.
$$
E(X) = \frac{E(U_1)}{p}, \ \ \ \ E(Y) = \frac{E(U_2)}{p}, 
$$
$$
V(X) = \frac{(1-p)(E(U_1))^2}{p^2} + \frac{V(U_1)}{p} \ \ \ \
V(Y) = \frac{(1-p)(E(V_1))^2}{p^2} + \frac{V(V_1)}{p}
$$
$$
\hbox{Cov}(X, Y) = \frac{1-p}{p^2}\ \ E(U_1)E(V_1).
$$
Some of the points are quite clear from the above moments expressions.  It is clear that as $p \rightarrow 0$, then the mean and
variance go to $\infty$.  It implies that the marginals become heavy tailed.  The correlation between $X$ and $Y$ is always 
positive and it goes to zero, as $p \rightarrow$ 1, and it goes to 1, as $p \rightarrow$ 0. The variance to mean ratio (VMR) for $X$ and $Y$ are
$$
\hbox{VMR}(X) = \frac{(1-p)E(U_1)}{p} + \hbox{VMR}(U_1) \ \ \ \ \hbox{and} \ \ \ \
\hbox{VMR}(Y) = \frac{(1-p)E(V_1)}{p} + \hbox{VMR}(V_1).
$$
It is clear that as $p \rightarrow$ 0, the marginals will be over dispersed, and for large $p$, the marginals can be 
under dispersed in the base line distributions are under dispersed.  Therefore, it is possible to have both over dispersed
and under dispersed marginals for the proposed bivariate distributions.  It is also possible to have two opposite behavior of 
the two marginals.  Now we will consider some special cases in the subsequent sections.

\section{\sc Two Special Cases}

\subsection{\sc Bivariate Poisson Geometric}

In this section it is assumed that $U_i$ follows $(\sim)$ a Poisson random variable with mean $\lambda_1$ 
(PO $(\lambda_1)$) and $V_i \sim$ PO $(\lambda_2)$.  We denote this new distribution as BPG $(\lambda_1, \lambda_2, p)$, 
and the marginals will be denoted by UPG $(\lambda_1, p)$ and UPG $(\lambda_2, p)$, respectively.
  
The joint PMF of $X$ and $Y$ for $m = 0, 1, \ldots$ and $n = 0, 1, \ldots$, can be obtained as follows:
\be
P \left (X = m, Y = n \right ) = 
C(\lambda_1+\lambda_2-\ln(1-p),m+n) \times \frac{p}{1-p} \times \frac{\lambda_1^m \lambda_2^n}{m! n!}.   \label{jpmfp}
\ee
Here for $a > 0$ and $j = 0, 1, 2, \ldots$,
$$
C(a,j) = \sum_{k=1}^{\infty} k^j \ e^{-ak}.
$$
The exact expressions of $C(a,j)$ for different values of $j$ can be obtained recursively, and it is provided in the 
Appendix A.  The marginal PMF of $X$ for $m = 0, 1, 2, \ldots$, and of $Y$ for $n = 0, 1, 2, \ldots$ can be obtained as,
\beanno
P \left (X = m \right ) & = & C(\lambda_1-\ln(1-p),m) \times \frac{p}{1-p} \times \frac{\lambda_1^m}{m!} \\ 
P \left (Y = n \right ) & = & C(\lambda_2-\ln(1-p),n) \times \frac{p}{1-p} \times \frac{\lambda_2^n}{n!},
\eeanno
respectively.

\begin{figure}[h]
\begin{center}
\includegraphics[height=4cm,width=6cm]{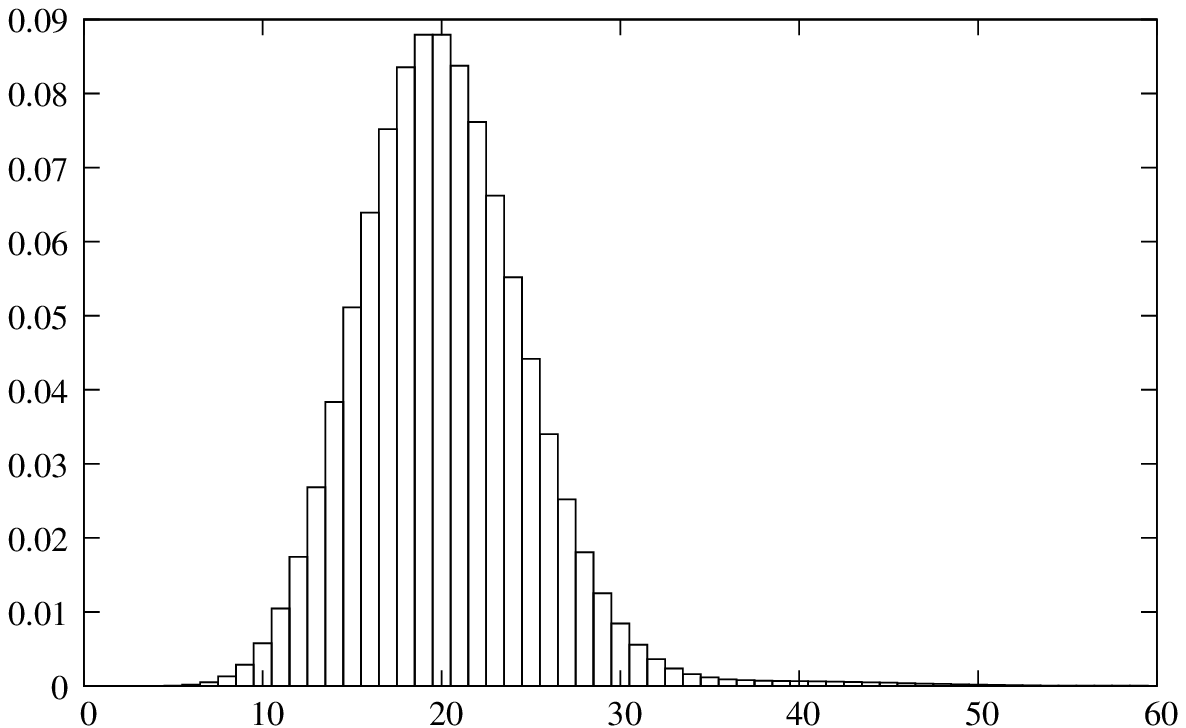}
\caption{The PMF of a univariate GDGE distribution when $\alpha$ = 1.5, $\theta$ = 0.5, $p = e^{-1.0}$.  \label{pmf-1}}
\includegraphics[height=4cm,width=6cm]{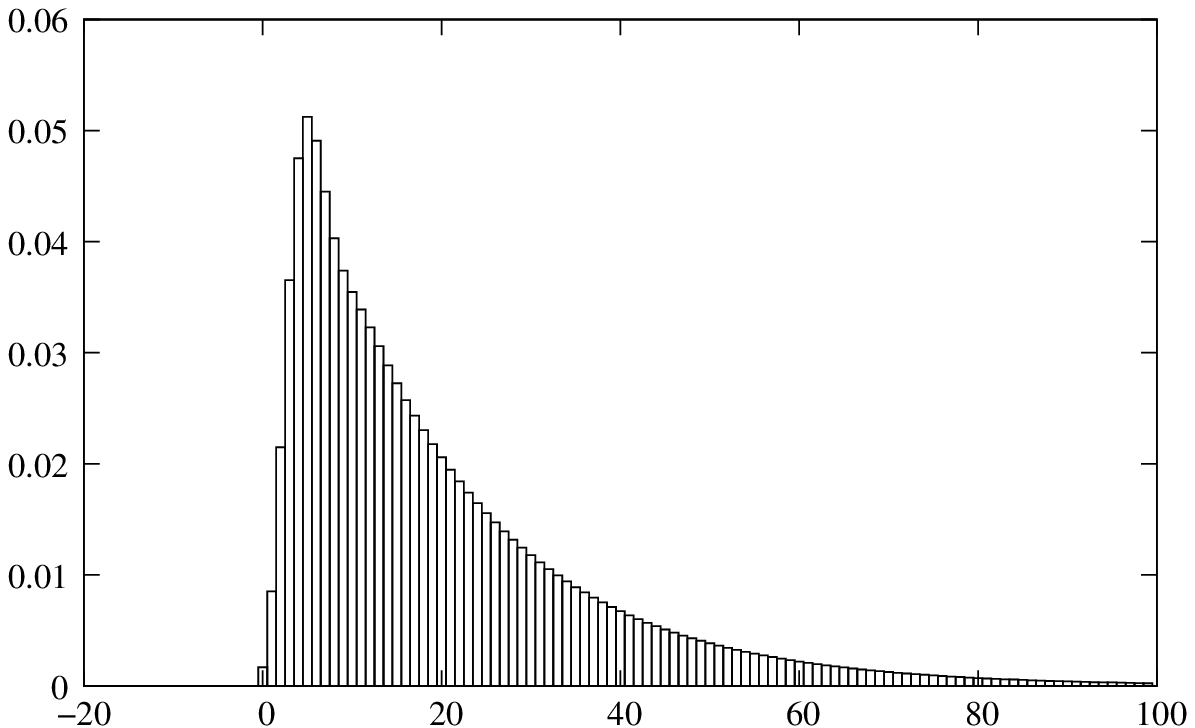}
\caption{The PMF of a univariate GDGE distribution when $\alpha$ = 1.5, $\theta$ = 0.01, $p = e^{-1.0}$.   \label{pmf-2}}
\includegraphics[height=4cm,width=6cm]{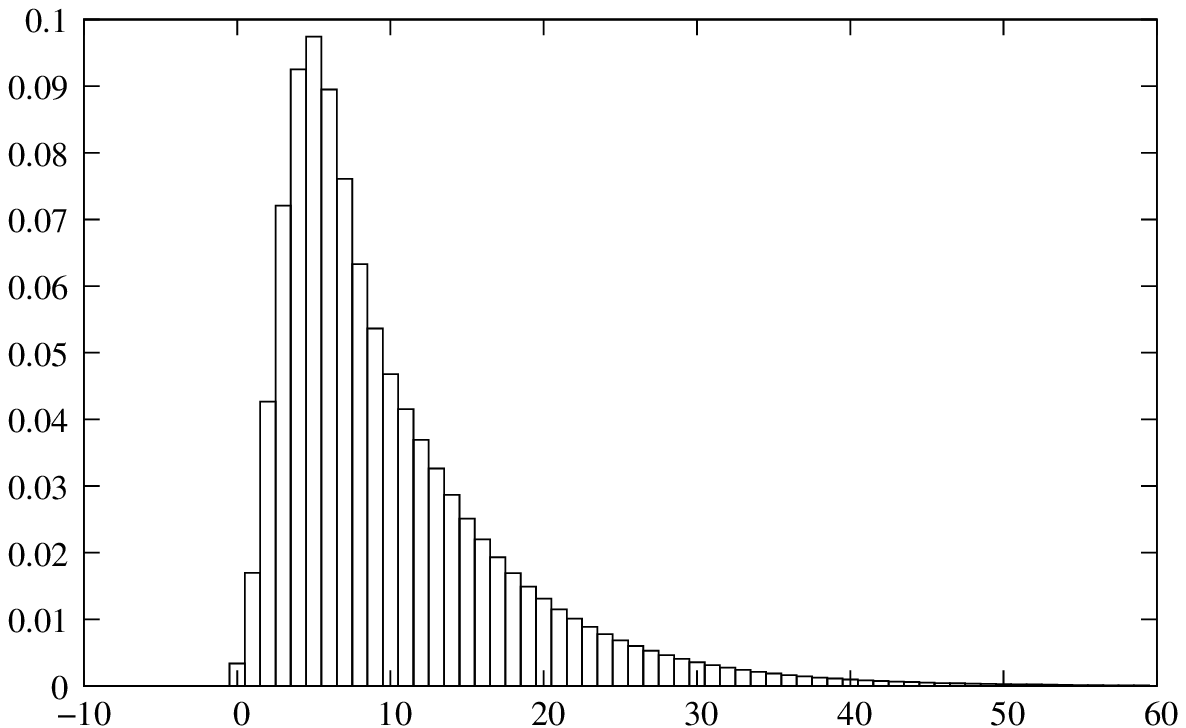}
\caption{The PMF of a univariate GDGE distribution when $\alpha$ = 1.5, $\theta$ = 0.01, $p = e^{-0.1}$.   \label{pmf-3}}
\includegraphics[height=4cm,width=6cm]{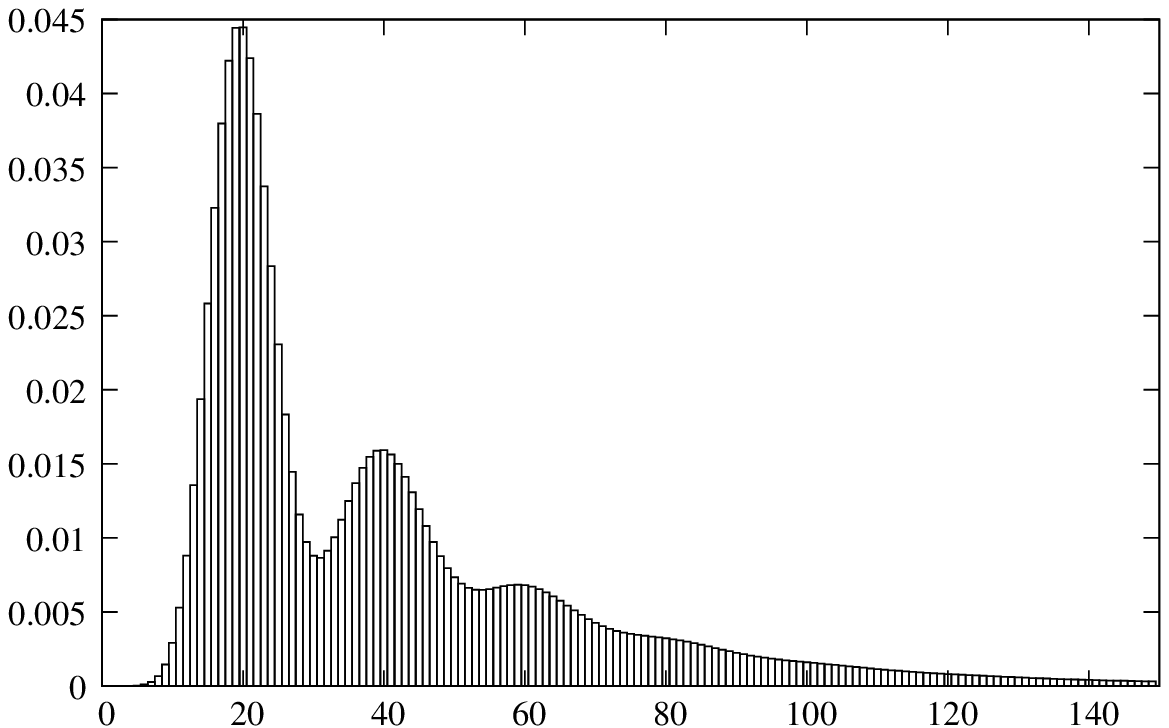}
\caption{The PMF of a univariate GDGE distribution when $\alpha$ = 1.5, $\theta$ = 0.01, $p = e^{-0.1}$.   \label{pmf-4}}
\end{center}
\end{figure}

$$
P(X = m| Y = n) = \frac{C(\lambda_1+\lambda_2-\ln(1-p),m+n)}{C(\lambda_1-\ln(1-p),m)} \times \frac{\lambda_1^m}{m!}.
$$
The joint moment generating function (MGF) of $X$ and $Y$ for $-\infty < s,t < \infty$ is
$$
\phi_{X,Y}(t, s) = E(e^{tX + s Y}) = \frac{p e^{\lambda_1(e^{t}-1)} e^{\lambda_2(e^{s}-1)}}{1 - (1-p)e^{\lambda_1(e^{t}-1)} e^{\lambda_2(e^{s}-1)}}.
$$
Hence the MGF of $X$ and $Y$ for $-\infty < t <\infty$, are  
$$
\phi_X(t) = E e^{tX} = \frac{p e^{\lambda_1(e^{t}-1)}}{1 - (1-p)e^{\lambda_1(e^{t}-1)}} \ \ \ \hbox{and} \ \ \ 
\phi_Y(t) = E e^{tY} = \frac{p e^{\lambda_2(e^{t}-1)}}{1 - (1-p)e^{\lambda_2(e^{t}-1)}} \ \ \ \hbox{and} \ \ \ 
$$
Different moments and correlation are
$$
E(X) = \frac{\lambda_1}{p}, \ \ \ \ E(Y) = \frac{\lambda_2}{p}, \ \ \ 
V(X) = \frac{\lambda_1^2}{p^2}, \ \ \ \
V(Y) = \frac{\lambda_2^2}{p^2}, \ \ \ \hbox{Corr}(X, Y) = (1-p).
$$
$$
\hbox{Cov}(X, Y) = \frac{1-p}{p^2}\ \ E(U_1)E(V_1).
$$
The VMR of $X$ and $Y$ are
$$
\hbox{VMR}(X) = \frac{\lambda_1}{p} \ \ \ \ \hbox{and} \ \ \ \
\hbox{VMR}(Y) = \frac{\lambda_2}{p}.
$$
Hence, depending on the parameters, the marginals can be over dispersed or under dispersed.  We have the following results regarding 
stochastic ordering of the Poisson geometric distribution.  The proofs are quite simple, hence
the details are avoided.

\noindent {\sc Result 1:} If $(X, Y) \sim$ BPG($\lambda_1$, $\lambda_2$, $p$) and $\lambda_1 > \lambda_2$, then $X$ is stochastically
larger than $Y$.

\noindent {\sc Result 2:} If $U \sim$ UPG$(\lambda, p_1)$ and $V \sim$ UPG$(\lambda, p_2)$, then for $p_1 < p_2$, $U$ is stochastically
larger than $V$.

\noindent {\sc Theorem 1:} If $(X, Y) \sim$ BPG($\lambda_1$, $\lambda_2$, $p$), then 
we have the following results.

\noindent (a) $X+Y \sim$ UPG$(\lambda_1+\lambda_2,p)$.

\noindent (b) $X$ given $X+Y$ follows a binomial distribution, i.e. 
$$
P(X = m|X+Y = n) = {n\choose{m}} \left ( \frac{\lambda_1}{\lambda_1+\lambda_2} \right )^m 
\left ( \frac{\lambda_2}{\lambda_1+\lambda_2} \right )^{n-m}; \ \ \ \ n = 0, 1, \ldots, m.
$$

\noindent {\sc Proof:} (a) can be obtained from the joint MGF.  (b) can be obtained using (a).

\noindent {\sc Theorem 2:} If $\{(X_i, Y_i); i = 1, 2, \ldots\}$ is a sequence i.i.d. random variables from \linebreak 
BPG($\lambda_1$, $\lambda_2$, $p$), $M$ is a geometric random variable with parameter $\delta$, for $0 < \delta < 1$,
and for $i = 1, \ldots$, it is independent of $(X_i, Y_i)$, then 
$$
\left ( \sum_{i=1}^M X_i, \sum_{i=1}^M Y_i \right ) \sim \hbox{BPG}(\lambda_1, \lambda_2, p \delta).
$$
\noindent {\sc Proof:} The proof mainly follows from the joint MGF. \qed

\noindent {\sc Theorem 3:} If $(X, Y) \sim$ BPG($\lambda_1$, $\lambda_2$, $p$), then $(X, Y)$ is infinitely divisible.

\noindent {\sc Proof:} Note that we need to prove that for any positive integer $n$, 
there exists i.i.d. random variables $\{(W^{(n)}_{1i}, W^{(n)}_{2i}); i = 1, 2, \ldots, n\}$, such that
$(W^{(n)}_{1i}, W^{(n)}_{2i})$ follows bivariate Poisson geometric random variables, and   
$$
(X, Y) \stackrel{d}{=} \left (\sum_{i=1}^n W^{(n)}_{1i}, \sum_{i=1}^n W^{(n)}_{2i} \right ).
$$
Here `$\stackrel{d}{=}$' means equal in distribution.  For any fixed $n$, let $r = 1/n$.  Consider a sequence of random 
i.i.d. random variables $Y_1, Y_2, \ldots$, such that $Y_i \sim$ PO$(r \lambda_1)$.  Similarly, consider another sequence of 
i.i.d. random variables $Z_1, Z_2, \ldots$, such that $Z_i \sim$ PO$(r \lambda_2)$.  Suppose $T$ is a negative binomial distribution 
NB$(r,p)$, with the following PMF
$$
P(T = k) = \frac{\Gamma(k+r)}{k! \Gamma (r)} p^r (1-p)^k; \ \ \  k = 0, 1, 2, \ldots.
$$
All the above random variables, namely $Y_i$'s $Z_i$'s and $T$ are independently distributed.  Consider the following bivariate 
random variable
$$
(W^{(n)}_{11}, W^{(n)}_{21}) \stackrel{d}{=} \left ( \sum_{i=1}^{1+nT} Y_i, \sum_{i=1}^{1+nT} Z_i \right ).
$$
The joint MGF of $(W^{(n)}_{11}, W^{(n)}_{21})$ can be obtained as 
\beanno
\phi_{W_{11}^{(n)}, W_{12}^{(n)}}(t,s) & = & E(e^{tW_{11}^{(n)} + s W_{21}^{(n)}}) = E_T \left (E \left (e^{tW_{11}^{(n)} + s W_{21}^{(n)}}|T \right ) \right )  \\
& = & E_T \left ( e^{r \lambda_1 (e^t-1)}e^{r \lambda_2 (e^2-1)(1+nT)} \right )  \\
& = & e^{r \lambda_1 (e^t-1)}e^{r \lambda_2 (e^2-1)} \left ( \frac{p}{1 - (1-p)e^{\lambda_1 (e^t-1)}e^{\lambda_2 (e^2-1)}} \right )^r   \\
& = & \left ( \frac{pe^{\lambda_1 (e^t-1)}e^{\lambda_2 (e^2-1)}}{1 - (1-p)e^{\lambda_1 (e^t-1)}e^{\lambda_2 (e^2-1)}} \right )^r.
\eeanno
Hence, the result is obtained.  \qed

The following result shows that the bivariate Poisson geometric distribution has an interesting decomposition.  It might have 
some independent interest also.

\noindent {\sc Theorem 4:} Let us assume that $(X, Y) \sim$ BPG($\lambda_1$, $\lambda_2$, $p$).  Suppose $Q \sim$ PO($\lambda$), 
where $\lambda = -\ln p$, and it is independent of $Z_i$'s, where $\{Z_i; i = 1, 2, \ldots\}$ is a sequence of i.i.d. random 
variables having logarithmic distribution with the probability mass function
\be
P(Z_1 = k) = \frac{(1-p)^k}{\lambda k}; \ \ \ k = 1, 2, \ldots.   \label{ld-pmf}    
\ee
Moreover, $\{W_{1i}; i = 1, 2, \ldots$\} is a sequence of i.i.d. random variables such that $W_{1i}|Z_i \sim$ PO($\lambda_1 Z_i$),
similarly, $\{W_{2i}; i = 1, 2, \ldots$\} is a sequence of i.i.d. random variables such that $W_{2i}|Z_i \sim$ PO($\lambda_2 Z_i$),
and conditionally they are independently distributed.  $R$ and $S$ are two independent random variables and they are independent of all 
the previous random variables, such that $R \sim$ PO$(\lambda_1)$, $S \sim$ PO$(\lambda_2)$.  They we have the following 
decomposition of $(X, Y)$
\be
(X, Y) \stackrel{d}{=} \left ( R + \sum_{i=1}^Q W_{1i}, S + \sum_{i=1}^Q W_{2i} \right ).   \label{th4}
\ee
\noindent {\sc Proof:} First observe that the probability generating function of $Q$ and $Z_1$ are
$$
E(t^Q) = e^{\lambda(t-1)}, \ \ \ t \in \mathbb{R} \ \ \ \ \hbox{and} \ \ \ \ 
E(t^{Z_1}) = \frac{\ln(1-(1-p)t)}{\ln p}, \ \ \ t < (1-p)^{-1}.
$$
The joint MGF of the right hand side of (\ref{th4}) can be written as
\beanno
\phi(u,v) & = & E \left (e^{u(R + \sum_{i=1}^Q W_{1i}) + v(S + \sum_{i=1}^Q W_{2i})} \right )  \\
& = & e^{\lambda_1(e^u-1)} e^{\lambda_2(e^v-1)}E \left (e^{u\sum_{i=1}^Q W_{1i} + v\sum_{i=1}^Q W_{2i}} \right )  \\
& = & e^{\lambda_1(e^u-1)} e^{\lambda_2(e^v-1)}E_Q \left [ E \left ( e^{u\sum_{i=1}^Q W_{1i} + v\sum_{i=1}^Q W_{2i}} \large | Q \right ) \right ]\\
& = & e^{\lambda_1(e^u-1)} e^{\lambda_2(e^v-1)}E_Q E \left [ E \left (e^{u\sum_{i=1}^Q W_{1i}} \large | Q, Z_1, Z_2, \ldots \right )
E \left ( e^{v\sum_{i=1}^Q W_{2i}} \large | Q, Z_1, Z_2, \ldots \right ) \right ]  \\
& = & e^{\lambda_1(e^u-1)} e^{\lambda_2(e^v-1)}E_Q E \left [ e^{\lambda_1(e^{ut}-1) \sum_{i=1}^Q Z_i} e^{\lambda_2(e^{vs}-1) \sum_{i=1}^Q Z_i} \right ] \\
& = & e^{\lambda_1(e^u-1)} e^{\lambda_2(e^v-1)}E_Q \left [ E \left [ e^{(\lambda_1(e^{ut}-1)+\lambda_2(e^{vs}-1)) Z_1} \right ] \right ]^Q  \\
& = & e^{\lambda_1(e^u-1)} e^{\lambda_2(e^v-1)}E_Q 
\left [ \frac{\ln (1 - (1-p) e^{(\lambda_1(e^{ut}-1)+\lambda_2(e^{vs}-1))})}{\ln p} \right ]^Q \\
& = & e^{\lambda_1(e^u-1)} e^{\lambda_2(e^v-1)} e^{\ln p - \ln (1 - (1-p) e^{(\lambda_1(e^{ut}-1)+\lambda_2(e^{vs}-1))})} \\
& = & \frac{pe^{\lambda_1(e^u-1)} e^{\lambda_2(e^v-1)}}{1-(1-p)e^{\lambda_1(e^u-1)} e^{\lambda_2(e^v-1)}}.
\eeanno

\subsection{\sc Bivariate Negative Binomial Geometric}

In this section we consider another special case when $U_1 \sim$ NB$(r_1, \theta_1)$ and $V_1 \sim$ NB$(r_2, \theta_2)$, where 
$r_1 > 0, r_2 > 0, 0 < \theta_1, \theta_2 < 1$.  Here NB$(r,\theta)$ means a negative binomial distribution with the PMF
\be
P(X = k) = \frac{\Gamma(k+r)}{k! \   \Gamma (r)} \theta^k (1-\theta)^r; \ \ \ k = 0, 1, 2, \ldots.
\ee
In this case we denote this bivariate distribution as BNBG$(r_1, \theta_1, r_2, \theta_2, p)$, and the marginals will be denoted by
UNBG$(r_1, \theta_1, p)$ and UNBG$(r_2, \theta_2, p)$, respectively.
The joint PMF of $X$ and $Y$ for $m = 0, 1, \ldots$ and $n = 0, 1, \ldots$, can be written as
\be
P \left (X = m, Y = n \right ) = 
D(r_1, r_2, \theta_1, \theta_2, m,n,p) \theta_1^m \theta_2^n \frac{p}{1-p},   \label{jpmfn}
\ee
where
$$
D(r_1, r_2, \theta_1, \theta_2, m,n,p) = \sum_{k=1}^{\infty}
\frac{\Gamma(m+kr_1)}{m! \   \Gamma (kr_1)} \times \frac{\Gamma(n+kr_2)}{n! \   \Gamma (kr_2)}
(1-p)^k (1-\theta_1)^{kr_1}(1-\theta_2)^{kr_2}.
$$
The marginal PMFs of $X$ and $Y$ can be obtained as
\beanno
P(X = m) & = & D_1(r_1,\theta_1,m,p) \theta_1^m \frac{p}{1-p}  \\
P(Y = n) & = & D_1(r_2,\theta_2,n,p) \theta_2^m \frac{p}{1-p},
\eeanno
where
$$
D_1(r,\theta,m,p) = \sum_{k=1}^{\infty} \frac{\Gamma(m+kr)}{m! \   \Gamma (kr)}
(1-\theta)^{kr} (1-p)^k.
$$
The joint MGF of $X$ and $Y$ for $-\infty < s,t < \infty$, is
$$
\phi_{X,Y}(t,s) = E(e^{t X + sY}) = \frac{p (1-\theta_1)^{r_1} (1-\theta_2)^{r_2}}{(1-\theta_1e^t)^{r_1}(1-\theta_2e^s)^{r_2} - 
(1-p)(1-\theta_1)^{r_1} (1-\theta_2)^{r_2}}.
$$
The marginal MGFs of $X$ and $Y$ can be obtained as
\be
\phi_X(t) = \frac{p (1-\theta_1)^{r_1}}{(1-\theta_1e^t)^{r_1} - 
(1-p)(1-\theta_1)^{r_1}}, \ \ \
\phi_Y(s) = \frac{p (1-\theta_2)^{r_2}}{(1-\theta_2e^s)^{r_2} - 
(1-p)(1-\theta_2)^{r_2}}.
\ee
Different moments and product moments of the BNBG distribution can be easily obtained and they are as follows.
$$
E(X) = \frac{\theta_1 r_1}{p(1-\theta_1)}, \ \ \ \ E(Y) = \frac{\theta_2 r_2}{p(1-\theta_2)}, 
$$
$$
V(X) = \frac{(1-p)\theta_1^2 r_1^2}{p^2(1-\theta_1)^2} + \frac{\theta_1 r_1}{p(1-\theta_1)^2} \ \ \ \
V(Y) = \frac{(1-p)\theta_2^2 r_2^2}{p^2(1-\theta_2)^2} + \frac{\theta_2 r_2}{p(1-\theta_2)^2} 
$$
$$
\hbox{Cov}(X, Y) = \frac{1-p}{p^2}\ \ \frac{\theta_1 r_1 \theta_2 r_2}{(1-\theta_1)(1-\theta_2)}.
$$
\noindent {\sc Theorem 5:} If $(X, Y) \sim$ BNBG($r_1$, $\theta$, $r_2$, $\theta$, $p$), then 
$X+Y \sim$ UNBG$(r_1+r_2,\theta,p)$.

\noindent {\sc Proof:} It can be obtained from the joint MGF.   \qed

\noindent {\sc Theorem 6:} If $\{(X_i, Y_i); i = 1, 2, \ldots\}$ is a sequence i.i.d. random variables from \linebreak 
BNBG($r_1$, $\theta_1$, $r_2$, $\theta_2$, $p$), and $M$ is a geometric random variable with parameter $\delta$, for 
$0 < \delta < 1$, then 
$$
\left ( \sum_{i=1}^M X_i, \sum_{i=1}^M Y_i \right ) \sim \hbox{BNBG}(r_1, \theta_1, r_2, \theta_2, p \delta).
$$
\noindent {\sc Proof:} The proof mainly follows from the joint MGF. \qed

\noindent {\sc Theorem 7:} If $(X, Y) \sim$ BNBG($r_1$, $\theta_1$, $r_2$, $\theta_2$, $p$), then $(X, Y)$ is infinitely divisible.

\noindent {\sc Proof:} The proof can be obtained along the same line of proof as of Theorem 3, the details are avoided.  \qed

\noindent {\sc Theorem 8:} Let us assume that $(X, Y) \sim$ BNBG($r_1$, $\theta_1$, $r_2$, $\theta_2$, $p$).  Suppose $Q \sim$ PO($\lambda$), 
where $\lambda = -\ln p$, and it is independent of $Z_i$'s, where $\{Z_i; i = 1, 2, \ldots\}$ is a sequence of i.i.d. random 
variables having logarithmic distribution with probability mass function as defined (\ref{ld-pmf}).
Moreover, $\{W_{1i}; i = 1, 2, \ldots$\} is a sequence of i.i.d. random variables such that $W_{1i}|Z_i \sim$ NB($r_1Z_i$, $\theta_1$),
similarly, $\{W_{2i}; i = 1, 2, \ldots$\} is a sequence of i.i.d. random variables such that $W_{2i}|Z_i \sim$ NB($r_2Z_i$, $\theta_2$),
and conditionally they are independently distributed.  $R$ and $S$ are two independent random variables and they are independent of all 
the previous random variables, such that $R \sim$ NB$(r_1, \theta_1)$, $S \sim$ NB$(r_2, \theta_2)$.  They we have the following 
decomposition of $(X, Y)$
\be
(X, Y) \stackrel{d}{=} \left ( R + \sum_{i=1}^Q W_{1i}, S + \sum_{i=1}^Q W_{2i} \right ).   \label{th8}
\ee
\noindent {\sc Proof:} The proof can be obtained similarly as the proof of Theorem 4.   \qed

\section{\sc Data Analysis}

\subsection{\sc Italian Football Data:} 

In this section we present the analysis of two data sets for illustrative purposes.  The first 
data set represents the Italian Series A football match score data between `ACF Firontina' ($X$) and `Juventus' ($Y$) during
1990 to 2005.  The data set is presented in Table \ref{football-data}.  We would like to use both BPG and BNBG to analyze this data set.  We have used the MMEs of 
the unknown parameters for both the models.  The sample means, variances and correlation are presented below.
$$
\mu_x = 1.3846, \ \ \ \ \sigma^2_x = 1.7751, \ \ \ \mu_y = 1.6923, \ \ \ \sigma^2_y = 2.2130, \ \ \ \ r_{x,y} = 0.1179.
$$
Based on the BPG model it can be easily seen that the MMEs of the unknown parameters and the associated bootstrap 
standard errors (reported within brackets) are
\beanno
\widetilde{p} & = & 1 - r_{x,y} = 0.8821 (\mp 0.0211), \\
\widetilde{\lambda}_1 & = & \mu_x (1 - r_{x,y}) = 1.2214 (\mp 0.1154) \\  
\widetilde{\lambda}_2 & = & \mu_y (1-r_{x,y}) = 1.4928 (\mp 0.1987).
\eeanno
Based on the BNBG model it can be easily seen that the MMEs of the unknown parameters and the associated bootstrap 
standard errors (reported within brackets) are
\beanno
\widetilde{p} & = & 1 - \frac{r_{x,y} \sigma_x \sigma_y}{\mu_x \mu_y} = 0.9003 (\mp 0.0114), \\
\widetilde{\theta}_1 & = & 1 - \mu_x/(\sigma^2_x - (1-\widetilde{p}) \mu_x^2) = 0.1259 (\mp 0.0013) \\ 
\widetilde{\theta}_2 & = & 1 - \mu_y/(\sigma^2_y - (1-\widetilde{p}) \mu_y^2) = 0.1221 (\mp 0.0019) \\  
\widetilde{r}_1 & = & \mu_x \widetilde{p}  (1-\widetilde{\theta}_1)/\widetilde{\theta}_1 = 8.6546 (\mp 0.7975) \\
\widetilde{r}_2 & = & \mu_y \widetilde{p}  (1-\widetilde{\theta}_2)/\widetilde{\theta}_2 = 10.9545 (\mp 1.1125) \\
\eeanno

\begin{table}[h]
\bc
\begin{tabular}{|l|c|c|l||c|c|}  \cline{1-6}

\hline
Obs. & ACF  & Juventus & Obs.   & ACF & Juventus  \\
     & Firontina  &    &    &    Firontina    &           \\
     &   ($X$)  & ($Y$) &  &  ($X$)  & ($Y$)    \\
   &   &   &   &   &   \\   \hline \hline
1 & 1 & 2 & 14 & 4 & 1    \\
2 & 0 & 0 & 15 & 4 & 4    \\
3 & 0 & 0 & 16 & 1 & 3    \\
4 & 2 & 2 & 17 & 1 & 3    \\
5 & 4 & 3 & 18 & 0 & 0    \\
6 & 0 & 1 & 19 & 1 & 0    \\
7 & 1 & 0 & 20 & 0 & 2    \\
8 & 3 & 2 & 21 & 3 & 0    \\
9 & 1 & 3 & 22 & 3 & 0    \\
10 & 2 & 0 & 23 & 1 & 2   \\
11 & 1 & 2 & 24 & 1 & 4   \\
12 & 2 & 3 & 25 & 0 & 2   \\
13 & 0 & 0 & 26 & 0 & 5   \\    \hline
\end{tabular}
\ec
\caption{UEFA Champion's League data \label{football-data}}
\end{table}

\subsection{\sc Seizure Data}

This data set represents weekly seizure data of 30 patients reported in Davis \cite{Davis:2002}.  Here the first column ($X$) and
the second column ($Y$) represent the number of seizures observed on each patient in the first week and second week, respectively, 
after admission to the hospital.  The data set is presented in Table \ref{seizure-data}.
\begin{table}[h]
\bc
\begin{tabular}{|l|c|c|l||c|c|}  \cline{1-6}

\hline
Obs. & Week-1 & Week-2 & Obs.   & Week-1 & Week-2  \\
     &   &    &    &        &           \\
     &   ($X$)  & ($Y$) &  &  ($X$)  & ($Y$)    \\
   &   &   &   &   &   \\   \hline \hline
1 & 5 & 0 & 16 & 1 & 0    \\
2 & 1 & 2 & 17 & 1 & 3    \\
3 & 1 & 4 & 18 & 0 & 2    \\
4 & 3 & 2 & 19 & 1 & 1    \\
5 & 3 & 1 & 20 & 2 & 1    \\
6 & 0 & 0 & 21 & 0 & 0    \\
7 & 1 & 0 & 22 & 2 & 1    \\
8 & 4 & 0 & 23 & 1 & 4    \\
9 & 0 & 0 & 24 & 1 & 0    \\
10 & 3 & 2 & 25 & 0 & 0   \\
11 & 3 & 0 & 26 & 2 & 2   \\
12 & 3 & 2 & 27 & 0 & 0   \\
13 & 1 & 0 & 28 & 1 & 1   \\    
14 & 3 & 2 & 29 & 6 & 0   \\ 
15 & 0 & 0 & 30 & 3 & 0   \\  \hline
\end{tabular}
\ec
\caption{Weekly seizure data on 30 patients \label{seizure-data}}
\end{table}
In this case also we would  like to use both BPG and BNBG to analyze this data set.  The sample means, variances and correlation 
are presented below.
$$
\mu_x = 1.5667, \ \ \ \ \sigma^2_x = 1.7789, \ \ \ \mu_y = 0.9667, \ \ \ \sigma^2_y = 1.4322, \ \ \ \ r_{x,y} = 0.0327.
$$
Based on the BPG model it can be easily seen that the MMEs of the unknown parameters and the associated bootstrap 
standard errors (reported within brackets) are
\beanno
\widetilde{p} & = & 1 - r_{x,y} = 0.9673 (\mp 0.0132), \\
\widetilde{\lambda}_1 & = & \mu_x (1 - r_{x,y}) = 1.5155 (\mp 0.1765) \\  
\widetilde{\lambda}_2 & = & \mu_y (1-r_{x,y}) = 0.9351 (\mp 0.0967).
\eeanno
Based on the BNBG model it can be easily seen that the MMEs of the unknown parameters and the associated bootstrap 
standard errors (reported within brackets) are
\beanno
\widetilde{p} & = & 1 - \frac{r_{x,y} \sigma_x \sigma_y}{\mu_x \mu_y} = 0.9655 (\mp 0.0178), \\
\widetilde{\theta}_1 & = & 1 - \mu_x/(\sigma^2_x - (1-\widetilde{p}) \mu_x^2) = 0.0753 (\mp 0.0004) \\ 
\widetilde{\theta}_2 & = & 1 - \mu_y/(\sigma^2_y - (1-\widetilde{p}) \mu_y^2) = 0.3094 (\mp 0.0065) \\  
\widetilde{r}_1 & = & \mu_x \widetilde{p}  (1-\widetilde{\theta}_1)/\widetilde{\theta}_1 = 18.5756 (\mp 1.2341) \\
\widetilde{r}_2 & = & \mu_y \widetilde{p}  (1-\widetilde{\theta}_2)/\widetilde{\theta}_2 = 2.0833 (\mp 0.0653) \\
\eeanno

\section{\sc Conclusions}

In this section we have proposed a very general bivariate discrete distributions which is a very flexible class of 
distributions.  Due to presence of an extra parameter, the proposed class of distributions is more flexible than the base 
distributuion functions.  It can take variety of shapes, and it can be both over dispersed as well as under dispersed depending 
on the parameters.  We have discussed several properties of the proposed class of distributions and consider two special 
cases, namely BPG and BNBG distributions.  It is observed that both BPG and BNBG are infinitely divisible and they have some
interesting physical interpretations also.  We have proposed to use the MMEs to estimate the unknown parameters and they can be 
obtained very conveniently and analyze two data sets for illustrative purposes.

Although, in this paper we have proposed the bivariate class of distributions, the method can be applied even for multivariate case
also.  Moreover, in this paper we have not developed any inference procedure based on the maximum likelihood approach.  It involves 
solving higher dimensional optmization problems.  Efficient algorithm is needed to compute the maximum likelihood estimators in this
case.  More work is needed along that direction.

\section*{\sc Appendix: Exact Expressions of $C(a,j)$}

Note that for
$$
C(a,0) = \sum_{k=1}^{\infty} e^{-a k} = \frac{e^{-a}}{1 - e^{-a}} = \frac{1}{e^a-1}.
$$
To compute $C(a,1)$, first observe that
$$
C(a,1) = \sum_{k=1}^{\infty} e^{-a k} k = \frac{1}{e^a} + \frac{2}{e^{2a}} + \frac{3}{e^{3a}} + \cdots 
$$
and
$$
e^a C(a,1) = 1 + \frac{2}{e^a} + \frac{3}{e^{2a}} + \frac{4}{e^{3a}} + \cdots.
$$
Hence
$$
(e^a - 1) C(a,1) = 1 + \frac{1}{e^a} + \frac{1}{e^{2a}} + \frac{1}{e^{3a}} + \cdots = \frac{e^a}{e^a - 1}.
$$
Therefore
$$
C(a,1) = \frac{e^a}{(e^a - 1)^2}.
$$
Now to compute $C(a,2)$, note that 
$$ 
C(a,2) = \sum_{k=0}^{\infty} e^{-a k} k^2 = \sum_{k=0}^{\infty} e^{-a k} k(k-1) + \sum_{k=0}^{\infty} e^{-a k} k.
$$ 
If we denote $\ds S =  \sum_{k=0}^{\infty} e^{-a k} k(k-1)$, then
$$
S = \frac{2 \cdot 1}{e^{2a}} + \frac{3 \cdot 2}{e^{3a}} +  \frac{4 \cdot 3}{e^{4a}} + \cdots 
$$
and
$$
e^a S = \frac{2 \cdot 1}{e^a} + \frac{3 \cdot 2}{e^{2a}} +  \frac{4 \cdot 3}{e^{3a}} + \cdots.
$$
Hence
$$
S(e^a-1) = \frac{2 \cdot 1}{e^a} + \frac{2 \cdot 2}{e^{2a}} +  \frac{2 \cdot 3}{e^{3a}} + \cdots = 2 \sum_{k=1}^{\infty} k e^{-ak}
= \frac{2 e^a}{(e^a-1)^2}.
$$
Therefore,
$$
C(a,2) = \frac{2 e^a}{(e^a-1)^3} + \frac{e^a}{(e^a-1)^2} = \frac{e^{2a} + e^a}{(e^a-1)^3}.
$$
We will use the following notations:
$$
S_0(a) = \sum_{k=0}^{\infty} k e^{-k a}, \ \ S_1(a) = \sum_{k=0}^{\infty} k(k-1) e^{-k a}, \ldots, S_m(a) = \sum_{k=0}^{\infty} k(k-1) \cdots (k-m) e^{-k a}.
$$
Then using the fact
$$
e^a S_m(a) = \sum_{k=0}^{\infty} k(k-1) \cdots (k-m) e^{-(k-1) a}.
$$
We can easily obtain the following relation
$$
S_m(a)(e^a-1) = (m+1) S_{m-1}(a).
$$
Further note that if we denote
$$
k^m = C_{0m} k(k-1) \cdots (k-m+1) + C_{1m} k(k-1) \cdots (k-m+1) + \cdots + C_{m-2,m} k (k-1) + C_{m-1,m} k,
$$
then $C_{0m}$, $C_{1m} \cdots, C_{m-1,m}$ can be obtained recursively from the following set of linear equations.  $C_{0m}$ = 1 
and
\beanno
-C_{0m} \sum_{1 \le i_1 \le m-1} i_1 + C_{1m} & = & 0  \\
C_{0m} \sum_{1 \le i_1 < i_2 \le m-1} i_1 i_2 - C_{1m} \sum_{1 \le i_1 \le m-2} i_1  + C_{2m} & = & 0 \\
- C_{0m} \sum_{1 \le i_1 < i_2 < i_3 \le m-1} i_1 i_2 i_3 + C_{1m} \sum_{1 \le i_1 < i_2 \le m-2} i_1 i_2  - C_{2m} \sum_{1 \le i_1 \le m-3} i_1  + C_{3m} & = & 0 \\
& \vdots &  \\
(-1)^{m-1} C_{0m} \prod_{i=1}^{m-1} i (-1)^{m-2} C_{1m} \prod_{i=1}^{m-2} i (-1)^{m-3} C_{2m} \prod_{i=1}^{m-3} i  + \cdots  - C_{m-2,m} + C_{m-1,m} & = & 0.  \eeanno
If we use the following notations for $n < m$;
$$
a_{nm} = \sum_{1 \le i_1 < i_2 < \ldots < i_n \le m} i_1 i_2 \cdots i_n, 
$$
$\ds a_{mm} = \prod_{i=1}^m i$, then clearly
$$
a_{n,m+1} = a_{n,m} + (m+1)a_{n-1,m},
$$
and we obtain
\beanno
C_{1m} & = & a_{1,m-1}  \\
C_{2m} & = & C_{1m} a_{1,m-2} - a_{2,m-1}  \\
C_{3m} & = & C_{2m} a_{1,m-3} - C_{1m} a_{2,m-2} + a_{3,m-1} \\
\vdots & = & \vdots \\
C_{m-1,m} & = & C_{m-2,m} a_{11} - C_{m-3,m} a_{2,2} + \ldots (-1)^{m-2} a_{m-1,m-1}.
\eeanno
Since we have
$$
C(a,m) = C_{0m} S_{m-1}(a) + C_{1m}S_{m-2}(a) + \ldots + C_{m-1,m} S_0(a),
$$
we can obtain recursively $C(a,m+1)$ from $C(a,m)$.

\end{document}